\documentclass[aps,prd,floatfix,nofootinbib]{revtex4} 
\usepackage{graphics}
\usepackage{graphicx}

\usepackage{latexsym}
\usepackage[cp866]{inputenc}
\usepackage[english]{babel}
\input epsf

\begin{document}

\title{\boldmath Coupled-channel influence on the $a_0(1700/1800)$ line shape}
\author{N. N. Achasov\,\footnote{achasov@math.nsc.ru}
and G. N. Shestakov\,\footnote{shestako@math.nsc.ru}}
\affiliation{Laboratory of Theoretical Physics, S. L. Sobolev
Institute for Mathematics, 630090, Novosibirsk, Russia}


\begin{abstract}
The current situation with the recently discovered $a_0(1700/1800)$
resonance is very paradoxical: it is believed that $a_0(1700/1800)$
must be strongly coupled with two vector mesons, but there is no
direct experimental confirmation of this yet. Based on the
assumption that the $a_0(1700/1800)$ is a state similar to the
four-quark state from the MIT bag, belonging to either the
$\underline{9}^*$ or the $\underline{36}^*$ $q^2\bar q^2$ multiplet,
we analyze the influence of the strong $a_0(1700/1800)$ coupling to
the vector channels $K^* \bar K^*$, $\rho\phi$, and $\rho \omega$ on
its line shape in the decay channels into pseudoscalar mesons $K\bar
K$, $\pi\eta$, and $\pi\eta'$. This effect depends on the location
of the resonance mass $m_{a_0}$ relative to the nominal thresholds
of vector channels. For example, if $m_{a_0}\approx 1700$ MeV, then
the influence turns out to be hidden in a fairly wide range of
coupling constants. On the whole, our analysis shows that, to
confirm the presence of the strong $a_0(1700/1800)$ coupling to the
vector channels, utterly required is the direct detection of the
decays $a_0(1700/1800)\to K^*\bar K^*$, $\rho\phi$, $\rho\omega$.
The appearance of even certain hints at the existence of these
decays would make it possible to fundamentally advance in
understanding the nature of the new $a_0$ state.
\end{abstract}

\maketitle


\section{Introduction}

Investigations of the $K\bar K$ and $\eta\pi$ mass spectra performed
recently by the BESIII, {\it BABAR}, and LHCb  Collaborations in
$D^+_s\to K^+K^-\pi^+$ \cite{Ab21},
$\gamma\gamma\to\eta_c\to\eta\pi^+\pi^-$ \cite{Le21}, $D^+_s\to
K^0_SK^0_S\pi^+$ \cite{Ab22a}, $D^+_s \to K^0_SK^+\pi^0$
\cite{Ab22b} and $B^+\to[\eta_c,\,\eta_c(2S),\,\chi_{ c1}]
K^+\to(K^0_SK^\pm\pi^\mp)K^+$ \cite{Aa23} indicate the existence of
a new scalar isovector resonance $a_0$ with a mass in the region of
1700$-$1800 MeV and a width of about 100 MeV. Below, we will denote
it conventionally as $a_0(1710)$ (or simply $a_0$). This state can
be a partner of the known isoscalar $f_0(1710)$ \cite{ PDG22}. The
BESIII \cite{Ab21,Ab22a,Ab22b} and {\it BABAR} \cite{Le21} data and
the earlier theoretical study \cite{Os09} stimulated discussion of
the nature of the $a_0(1710)$ state and its possible manifestations
in other reactions as well as construction of the models for
description of the experimentally observed two-body mass spectra
taking into account the $a_0(1710)$ contribution
\cite{Os22a,Wa22,Zh22a,Gu22,Zh22b,Os23a,Os23b,Wa23,Wan23,Di23}. In
most of these works, the $a_0(1710)$ is considered as a state
dynamically generated by interactions between the vector mesons,
including their coupling with the channels of the pseudoscalar
mesons in the framework of the coupled-channel approach $K^*\bar
K^*$, $\rho\phi$, $\rho\omega$, $K\bar K$, and $\pi\eta$ \cite{Os09,
Os22a,Wa22,Os23b}. For the present, the $a_0(1710)$ state was
observed only in the decay channels into $K\bar K$ and $\pi\eta$
that are not suppressed by the phase space. Note that the
experimental data on the mass and total width of the $a_0(1710)$ in
these channels \cite{Ab21,Le21,Ab22a, Ab22b,Aa23} were obtained
within the framework of the isobar model in which the usual
relativistic Breit-Wigner formulas were used to describe the
resonant contributions. Of course, it would be interesting to find
direct evidence confirming the strong $a_0(1710)$ coupling to the
decay channels into two vector mesons.

Recall that the activity in the sector of scalar mesons in the
region of 1800 MeV in the channels $K^*\bar K^*$, $\rho\phi$,
$\rho\omega$, $K\bar K$, $\pi\eta$, and $\pi\eta'$ with isospin
$I=1$ and in similar channels with $I=0$ was predicted 46 years ago
by Jaffe \cite{Ja77} within  the MIT bag model which
phenomenologically takes into account quark confinement. This
activity owes to the four-quark scalar states
$C^s_\pi(\underline{9}^*)$ and $C_\pi(\underline{36}^*)$ with $I=1$
(analogs of $a_0$) and also the $C^s(\underline{9}^*)$ and
$C^0(\underline{36}^*)$ with $I=0$ (analogs of $f_0$) belonging to
the $\underline{9}^*$ and $\underline{36}^*$ four-quark multiplets
\cite{Ja77}. The Jaffe model also predicts the expansion of the $q^2
\bar q^2$ state wave functions in terms of the $(q\bar q)(q\bar q)$
states $PP$, $VV$, $\underline{P}\cdot \underline{P}$, and
$\underline{V }\cdot \underline{V}$, where symbols $P$
($\underline{P}$) and $V$ ($\underline{V}$) denote colorless (color)
pseudoscalar and colorless (color) vector $q\bar q$ mesons,
respectively. The correct $q^2 \bar q^2\to(q\bar q)(q\bar q)$
recoupling coefficients for the $J=0$ four-quark bag states were
obtained in Ref. \cite{Wo80} (see also Refs. \cite{Ac82a, Ac82b}).
These coefficients make it possible to form the rough idea of the
relative strength of the coupling between the four-quark states and
decay channels into pairs of pseudoscalar and vector mesons.

In the present work, we consider two scenarios in which the
$a_0(1710)$ resonance is strongly coupled to the decay channels into
vector mesons ($VV$), and analyze the influence of this coupling on
the line shape of the $a_0(1710)$ in its decay channels into
pseudoscalar mesons ($PP$). In the first scenario, the $a_0(1710)$
is treated as a four-quark state containing a hidden $s\bar s$ pair
and having the Okubo-Zweig-Iizuka (OZI)-superallowed coupling
\cite{Ja77} to $K^*\bar K^*$, $\rho\phi$, $K\bar K$, and $\pi\eta_s$
($\eta_s$ is shorthand for $s\bar s$). Such a state can be
associated with the state $C^s_\pi(\underline{9}^*)$ in the Jaffe
model \cite{Ja77}. According to the second scenario, the $a_0(1710)$
is a four-quark state without strange quarks having the
OZI-superallowed coupling to $\rho\omega$ and $\pi\eta_0$ [$\eta_0$
denotes $(u\bar u+d\bar d)/\sqrt{2}$]. Its analog in the model
\cite{Ja77} is the state $C_\pi(\underline{36}^*)$. The paper is
organized as follows. Section II contains the necessary formulas for
describing the solitary $a_0(1710)$ resonance. Section III analyzes
in detail the influence of coupled channels on the shape of the $PP$
and $VV$ mass spectra in the $a_0(1710)$ decays depending on the
location of the resonance mass relative to the nominal thresholds of
the vector channels and on the values of the coupling constants. In
many cases, the influence of the strong coupling of the $a_0(1710)$
to the vector channels on its line shape in the decay channels into
pseudoscalar mesons turns out to be hidden or difficult to
distinguish. Therefore, the decisive confirmation of the existence
of a strong $a_0(1710)$ coupling  to the vector channels would be
the direct detection of the decays $a_0(1710)\to K^*\bar K^*$,
$\rho\phi$, $\rho\omega$. Section IV summarizes our conclusions.


\section{\boldmath Solitary $a_0(1710)$ resonance}

Consider the solitary $a^+_0(1710)$ resonance coupled to the decay
channels $ab=K^{*+}\bar K^{*0}$, $\rho^+\phi$, $\rho^+\omega$,
$K^+\bar K^0$, $\pi^+\eta$, and $\pi^+\eta'$ (in the following, we
will indicate the charge of the $a_0(1710)$ only if necessary). The
$a_0(1710)$ propagator taking into account the finite width
corrections has the form \cite{GS68,BM73,Fl76,Ac80a,Ac80b}
\begin{eqnarray}\label{Eq1}
\frac{1}{D_{a_0}(s)}=\frac{1}{m^2_{a_0}-s+\sum_{ab}[\mbox{Re}
\Pi^{ab}_{a_0}(m^2_{a_0})-\Pi^{ab}_{a_0}(s)]},\end{eqnarray} where
$s$ is the square of the invariant mass of the virtual $a_0(1710)$
state, $m_{a_0}$ is a mass of the $a_0(1710)$, and $\Pi^{ab}_{a_0}
(s)$ is the matrix element of the $a_0(1710)$ polarization operator
corresponding to the contribution of the $ab$ intermediate state.
The energy-dependent total width of the $a_0(1710)$ is given by
\begin{eqnarray}\label{Eq1a} \Gamma^{
\scriptsize\mbox{tot}}_{a_0}(s)=-\mbox{Im}\,D_{a_0} (s)/\sqrt{s}=
\sum_{ab}\mbox{Im}\,\Pi^{ab}_{a_0}(s)/\sqrt{s}\,.\end{eqnarray} The
masses of particles in $\pi$, $K$, and $K^*$ isotopic multiplets are
putted to be equal to the masses of $\pi^+$, $K^+$, and $K^{*+}$
mesons, respectively. In the case of the $a_0(1710)$ decay into
pairs of pseudoscalar mesons, the imaginary part of
$\Pi^{ab}_{a_0}(s)$, which is nonzero for $s>(m_a+m_b)^2$, has the
form \begin{eqnarray}\label{Eq2}
\mbox{Im}\,\Pi^{ab}_{a_0}(s)=\sqrt{s}\Gamma_{a_0\to ab}(s)=
\frac{g^2_{a_0 ab}}{16\pi}\rho_{ab}(s),
\end{eqnarray} where $g_{a_0ab}$ is the coupling constant of the
$a_0(1710)$ to the $ab$ channel,\, $\rho_{ab}(s)=\sqrt{s-m_{ab}^{
(+)\,2}}\,\sqrt{ s-m_{ab}^{(-)\,2}}/s$, and $m_{ab}^{(\pm)}
$\,=\,$m_a \pm m_b$. In so doing, $\Pi^{ab}_{a_0}(s)$ is given by
the once subtracted dispersion integral corresponding to the
one-loop $S$-wave Feynman diagram with particles $ab$ ($K\bar K$,
$\pi\eta$, $\pi\eta'$) in the intermediate state:
\begin{eqnarray}\label{Eq3} \Pi^{ab}_{a_0}(s)=\frac{s}{\pi}
\int\limits^\infty_{m_{ab}^{(+)\,2}}\frac{\sqrt{s'}\Gamma_{a_0\to
ab}(s')\,ds'}{\,s'(s'-s-i\varepsilon)}\,.\end{eqnarray} For
$s>m_{ab}^{(+)\,2}$,
\begin{eqnarray}\label{Eq4} \Pi^{ab}_{a_0}(s)=\frac{g^2_{a_0 ab}}
{16\pi}\left[L_{ab}(s)+ \rho_{ab}(s) \left(i-\frac{1}{\pi}\,
\ln\frac{\sqrt{s-m_{ab}^{(-) \,2}}+\sqrt{s-m_{ab}^{(+)\,2}}}{
\sqrt{s-m_{ab}^{(-)\,2}}-\sqrt{s -m_{ab}^{(+)\,2}}}\right)\right],
\\ \label{Eq5} L_{ab}(s)=\frac{1}{\pi}\left[1+\left(\frac{m_{ab}^{
(+)\,2}+ m_{ab}^{(-)\,2}}{2m_{ab}^{(+)}m_{ab}^{(-)}}-
\frac{m_{ab}^{(+)} m_{ab}^{(-)}}{s}\right)\ln\frac{m_a}{m_b}
\right].\ \end{eqnarray} For $m_{ab}^{(-)\,2}<s<m_{ab}^{(+)\,2}$,
\begin{eqnarray}\label{Eq6}
\Pi^{ab}_{a_0}(s)=\frac{g^2_{a_0 ab}} {16\pi}\left[L_{ab}(s)-\rho_{
ab}(s)\left(1-\frac{2}{\pi} \arctan\frac{\sqrt{ m_{ab}^{(+)\,2}-s
}}{\sqrt{s-m_{ab}^{(-)\,2}}} \right)\right],\end{eqnarray} where
$\rho_{ab}(s)=\sqrt{ m_{ab}^{(+)\,2}-s}\,\sqrt{s-m_{ab}^{
(-)\,2}}\,/s$. The region $s<m_{ab}^{(-)\,2}$ will not be
considered.

If the vector mesons were stable, we can use Eqs.
(\ref{Eq2})$-$(\ref{Eq6}) for an estimate of the contributions of
vector intermediate states. Let us write the $S$-wave amplitude of
the $a_0(1710)$ decay into a pair of vector mesons as
\begin{eqnarray}\label{Eq7} \mathcal{A}^{{\scriptsize
S-\mbox{wave}}}_{a_0\to ab}= \frac{g_{a_0ab}}{\sqrt{3}}\,
\epsilon^*_{a\mu}\epsilon^{*\mu}_b, \end{eqnarray} where
$\epsilon_a$ ($\epsilon_b$) is the polarization four-vector of the
$a$ ($b$) vector meson and $g_{a_0ab}$ is the corresponding coupling
constant. Then the calculation of the width of the $a_0\to VV$ decay
near its threshold leads in the nonrelativistic approximation
exactly to Eq. (\ref{Eq2}). The next approximation is to use Eq. (3)
for all $s$ above the $ab$ threshold in order to calculate
$\Pi^{ab}_{a_0}(s)$ according to Eq. (4) and as a result to have the
expressions (5)$-$(7) for contributions of the vector channels.
Below, we illustrate the specific differences between this
hypothetical variant and the variant that takes into account the
finite widths of the vector mesons.

Because of the limited phase spaces of the $VV$ states near their
nominal thresholds, i.e., at $\sqrt{s}\approx m_\rho+m_\phi
\approx1.795$ GeV, $\sqrt{s}\approx 2m_{K^*}\approx1.791$ GeV, and
$\sqrt{s} \approx m_\rho+m_\omega \approx1.558$ GeV, the finite
widths of the $V$ mesons must be taken into account. We will do this
for the $K^*$ and $\rho$ mesons, while the $\phi$  and $\omega$
mesons will be considered in the zero-width approximation
\cite{PDG22}. Let us start with the $a^+_0\to K^{*+}\bar
K^{*0}\to(K\pi)^+(\bar K\pi)^0$ decay. Isotopic weights of its four
charged modes are listed in Table I.
\begin{center}\begin{table}  [!ht] \caption{Isotopic weights of the
 $a^+_0\to K^{*+}\bar K^{*0}\to(K\pi)^+(\bar K\pi)^0$ decay modes.}
\begin{tabular}{ c c c c c } \hline\hline
  \ Decay mode of\  & (a) & (b) & (c) & (d) \\
  \ \ $a^+_0\to K^{*+}\bar K^{*0}$\  & \ \ $K^0\pi^+K^-\pi^+$ & \ \ $K^0\pi^+
  \bar K^0\pi^0$ & \ \ $K^+\pi^0K^-\pi^+$ & \ \ $K^+\pi^0\bar K^0\pi^0$\ \ \\
  \hline \ \ Isotopic weight\ \ & 4/9 & 2/9 & 2/9 & 1/9  \\
  \hline\hline\end{tabular}\end{table}\end{center}

The $a_0$ decay into channel (b) is described by one amplitude
$a^+_0\to K^{*+}\bar K^{*0}\to(K^0\pi^+)(\bar K^0\pi^0)$. The same
applies to the $a_0$ decay into channel (c). The $a_0$ decay into
channel (a) is described by two amplitudes differing by the
permutation of two identical $\pi^+$ mesons: $a^+_0\to K^{*+}\bar
K^{*0}\to[(K^0\pi^+_1)(\bar K^-\pi^+_2)+ (K^0\pi^+_2)(\bar
K^-\pi^+_1)]$. Their contribution enters into the $a_0$ decay width
with a factor of $1/2!$. The same applies to the $a_0$ decay into
channel (d). The modulus squared of each charged amplitude gives
(without taking its isotopic weight into account) an equal
contribution to the $a_0$ width. As is seen from  Table I, the sum
of isotopic weights of all charged $a_0$ decay modes is normalized
to 1 and the weight of the interference contribution originating
from channels (a) and (d) is equal to 5/9. We note that calculations
of the widths of the resonances decaying into $VV$ channels may be
found, for example, in Refs. \cite{Ac82b,Po86,Bu87,Ac91} (see also
references therein).

We denote by $s_1$ and $s_2$ the squares of the invariant masses of
the virtual $K^{*+}$ and $\bar K^{*0}$ mesons, respectively, and
weigh the $S$-wave two-body invariant phase space
\begin{eqnarray}\label{Eq8} \rho(s,s_1,s_2)=\sqrt{
s^2-2s(s_1+s_2)+(s_1-s_2)^2}/s \end{eqnarray} for the $K^{*+}\bar
K^{*0}$ pair with the resonant $K^{*+}$ and $\bar K^{*0}$
Breit-Wigner distributions (which we assume to be the same)
\begin{eqnarray}\label{Eq9}
\bar{\rho}_{K^{*+}\bar K^{*0}}(s)=\frac{1+\frac{5}{9}C(s)}{\pi^2}
\int\limits^{(\sqrt{s}-m_K-m_\pi)^2}_{(m_K+m_\pi)^2} \frac{\sqrt{
s_1}\Gamma_{K^*}(s_1)}{|D_{K^*}(s_1)|^2}\, ds_1 \int\limits^{(
\sqrt{s}-\sqrt{s_1})^2}_{(m_K+m_\pi)^2}\frac{\sqrt{s_2}
\Gamma_{K^*}(s_2)}{|D_{K^*}(s_2)|^2}\,\rho(s,s_1,s_2)\, ds_2.
\end{eqnarray} Here, \begin{eqnarray}\label{Eq10}
D_{K^*}(s_j)=m^2_{K^*}-s_j-i\sqrt{ s_j}\,\Gamma_{K^*}(s_j)\,,
\end{eqnarray} \begin{eqnarray}\label{Eq11}
\sqrt{s_j}\,\Gamma_{K^*}(s_j)=m_{K^*}\Gamma_{K^*}(m^2_{K^*})
\frac{m_{K^*}}{\sqrt{s_j}}\left(\frac{q_{K\pi}(s_j)}{q_{K\pi}
(m^2_{K^*})}\right)^3 \frac{ 1+q^2_{K\pi}(m^2)r^2_{K^*}}{
1+q^2_{K\pi}(s_j)r^2_{K^*}}\,,
\end{eqnarray} $q_{K\pi}(s_j)=\sqrt{s_j}\rho(s_j,m^2_K,m^2_\pi)/2$,
$j=1,2$; $m_{K^*}=0.8955$ GeV, $\Gamma_{K^*}(m^2_{K^*})=0.05$ GeV,
and $r_{K^*}=3$ GeV$^{-1}$ \cite{PDG22}. The function $C(s)$ in Eq.
(\ref{Eq9}) describes the relative contribution of interference
between diagrams differing by the permutation of two identical $\pi$
mesons in the decay modes (a) and (d) in Table I. $C(s)$ is a smooth
function of $s$, $0<C(s)<1$; it tends to zero as $s$ increases and
also for $\Gamma_{K^*}(m^2_{K^*})\to0$. The available estimate of
the interference contribution in the reaction
$\gamma\gamma\to\rho^0\rho^0\to\pi^+\pi^-\pi^+\pi^-$ \cite{Ac91}
shows that the quantity $\frac{5}{9}C(s)$ is certainly less than
0.1, and we neglect this contribution. So, taking into account the
finite width of the $K^*$, we get
\begin{eqnarray}\label{Eq12}
\mbox{Im}\,\Pi^{K^{*+}\bar K^{*0}}_{a_0}(s)=\sqrt{s}\Gamma_{a_0\to
K^{*+}\bar K^{*0}}(s)=\frac{g^2_{a_0 K^{*+}\bar K^{*0}}}{16\pi}
\bar{\rho}_{K^{*+}\bar K^{*0}}(s)\,,\end{eqnarray}
\begin{eqnarray}\label{Eq13}
\Pi^{K^{*+}\bar K^{*0}}_{a_0}(s)=\frac{s\,g^2_{a_0 K^{*+}\bar K^{*0}
}}{16\pi^2} \int\limits^\infty_{(2m_K+2m_\pi)^2}\frac{\bar{\rho
}_{K^{*+}\bar K^{*0}}(s')\,ds'}{ \,s'(s'-s-i\varepsilon)}\,.
\end{eqnarray}
Similarly, taking into account the finite width of the $\rho$ meson
in the decays $a^+_0\to\rho^+(\phi/\omega)\to\pi^+\pi^0(\phi/\omega
)$, we have
\begin{eqnarray}\label{Eq14}
\bar{\rho}_{\rho^+(\phi/\omega)}(s)=\frac{1}{\pi}\int\limits^{
(\sqrt{s}-m_{\phi/\omega})^2}_{4m^2_\pi}\frac{\sqrt{ s_1}\Gamma_{
\rho}(s_1)}{|D_{\rho}(s_1)|^2} \,\rho(s,s_1,m^2_{\phi/\omega})\,
ds_1\,,\end{eqnarray}
\begin{eqnarray}\label{Eq12a}
\mbox{Im}\,\Pi^{\rho^+(\phi/\omega)}_{a_0}(s)=\sqrt{s}\Gamma_{a_0\to
\rho^+(\phi/\omega)}(s)=\frac{g^2_{a_0\rho^+(\phi/\omega)}}{16\pi}
\bar{\rho}_{\rho^+(\phi/\omega)}(s)\,,\end{eqnarray}
\begin{eqnarray}\label{Eq13a}
\Pi^{\rho^+(\phi/\omega)}_{a_0}(s)=\frac{s\,g^2_{a_0\rho^+(\phi/
\omega)}}{16\pi^2}\int\limits^\infty_{(m_{\phi/\omega}+2m_\pi
)^2}\frac{\bar{\rho }_{\rho^+(\phi/\omega)}(s')\,ds'}{\,s'
(s'-s-i\varepsilon)}\,.
\end{eqnarray}
The functions $D_{\rho}(s_1)$ and $\sqrt{s_1}\Gamma_{ \rho}(s_1)$
are obtained from Eqs. (\ref{Eq10}) and (\ref{Eq11}), respectively,
by using the obvious changing of indexes. Here we put $m_\phi=1.019
61$ GeV, $m_\omega=0.78265$ GeV, $m_\rho=0.77526$ GeV, $\Gamma_\rho
(m^2_\rho )=0.1491$ GeV \cite{PDG22}, and $r_\rho=1.5$ GeV$^{-1}$
\cite{Aa22}.

Let us now discuss the relations between the coupling constants
$g_{a_0ab}$. According to these relations, one can judge to some
extent about the nature of the decaying state. In the MIT bag model,
the flavor structure of the wave functions of the four-quark scalars
$C^s_\pi(\underline{9}^*)$ and $C_\pi(\underline{36}^*)$ with masses
about 1800 MeV  has the form \cite{Ja77,Wo80}
\begin{eqnarray}\label{Eq14} C^s_\pi(\underline{9}^*)=
-0.177\,\left(-\frac{1}{\sqrt{2}}\,K\bar K-\frac{1}{\sqrt{2}}
\,\eta_s\pi\right)+0.644\,\left(-\frac{1}{\sqrt{2}}\,K^*\bar
K^*-\frac{1}{\sqrt{2}}\,\phi\rho\right)+\cdot\cdot\cdot\,,
\end{eqnarray} \begin{eqnarray}\label{Eq15} C_\pi(\underline{36}^*)=
0.041\,(\pi\eta_0)+0.743\,(\rho\omega)+\cdot\cdot\cdot\,.\end{eqnarray}
Here $\eta_0=(u\bar u+d\bar d)/\sqrt{2}$ and $\eta_s=s\bar s$ are
the linear combinations of the physical states $\eta$ and $\eta'$:
$\eta_0=\eta\sin(\theta_i-\theta_p)+\eta'\cos(\theta_i-\theta_p)$
and $\eta_s=s\bar s=\eta'\sin(\theta_i-\theta_p) -\eta\cos(
\theta_i-\theta_p)$, where $\theta_i =35.3^\circ$ is the so-called
``ideal'' mixing angle and $\theta_p= -11.3^\circ$ is the mixing
angle in the nonet of the light pseudoscalar mesons \cite{PDG22}.

The coefficients in front of the expressions in parentheses in the
right-hand sides of Eqs. (\ref{Eq14}) and (\ref{Eq15}) are taken
from Table II; the dots imply that the wave functions of the
$C^s_\pi(\underline{9}^*)$ and $C_\pi(\underline{36}^*)$ states
involve the contributions from $q\bar q$ pairs with hidden color
$\underline{P}\cdot\underline{P}$ and $\underline{V}\cdot
\underline{V}$ (see Table II).
  \begin{center}\begin{table}  [!ht] \caption{Recoupling coefficients for
  $J^P=0^+$ $q^2\bar q^2$ mesons into two $q\bar q$ mesons \cite{Wo80}.}
  \begin{tabular}{ c l l l l } \hline\hline
  \,\,\,\,Flavor\,\, & $\quad PP\quad $  & $\quad VV\quad $ & $\quad \underline{P}\cdot\underline{P}\quad $ &
  $\quad \underline{V}\cdot\underline{V}\quad $ \\ \hline
  \,\,\,\,$\underline{9}^*$\,\, & $-0.177$ & $\quad 0.644$ & $\quad 0.623$ & $\quad 0.407$ \\
  \,\,\,\,$\underline{36}^*$\,\,\,\, & \ \ \,$0.041\,\, $ & $\quad 0.743$ & \ $ -0.646$\, & \ $ -0.169$\, \\
  \hline\hline \end{tabular}\end{table}\end{center}
Thus, if we identify $a_0(1710)$, for example, with the state
$C^s_\pi(\underline{9}^*)$, then we will have the following coupling
constants $a_0(1710)$ with $PP$ and $VV$ channels: $g_{a^+_0K^+\bar
K^0}=0.177\,\frac{1}{\sqrt{2}}\,g_0$, $g_{a^+_0\eta \pi^+}=-0.177\,
\frac{1}{\sqrt{2}}\,g_0\cos(\theta_i-\theta_p)$, $g_{a^+_0\eta'\pi^+
}=0.177\,\frac{1}{\sqrt{2}}\,g_0\sin(\theta_i -\theta_p)$, $g_{a^+_0
K^{*+}\bar K^{*0}}=g_{a^+_0\rho^+\phi}= -0.644\,\frac{1}{\sqrt{2}}
\,g_0$, where the universal coupling constant $g_0$ describes the
OZI-superallowed decays of the $q^2\bar q^2$ mesons into two $q\bar
q$ mesons \cite{Ja77}. We will not strictly adhere to the relations
between the $PP$ and $VV $ components of the wave functions that
follow from Table II.

As for the masses of $q^2\bar q^2$ states in the MIT bag model, they
were quoted in Ref. \cite{Ja77} to the nearest 50 MeV. The modern
values of masses of the light scalar mesons $f_0(500)$, $K^*_0(700)
$, $a_0(980)$, and $f_0(980)$ \cite{PDG22}, which are candidates for
the four-quark states \cite{Ja77,Ac89,Ac98,Ac03a}, indicate a shift
about 100 MeV toward lower masses relative to the predictions of the
$q^2\bar q^2$ model \cite{Ja77}. In this regard, any value for the
mass of the $a_0(1710)$ in the region 1700$-$1800 MeV seems quite
probable within the $q^2\bar q^2$ model.

Let us note  two points. First, the $q^2\bar q^2$ MIT bag model
predicts a strong coupling of the $a_0(1710)$ to two vector mesons,
as is seen from Eqs. (18) and (19). For example, the ratios $g^2_{
a^+_0 K^{*+}\bar K^{*0}}/g^2_{a^+_0K^+\bar K^0}$ and $g^2_{a^+_0
\rho^+\phi}/g^2_{a^+_0K^+\bar K^0}$  for the $a_0(1710)=C^s_\pi(
\underline{9}^*)$ are approximately 13.2. If these ratios are of the
order of 1, then the searches for the $a_0(1710)\to VV$ decays are
hopeless. Second, the $q^2\bar q^2$ model predicts the existence of
the isoscalar partner of the $a_0(1710)$ with a close or even
degenerate mass. The known $f_0(1710)$ state \cite{PDG22} is one of
the possible candidates to this role. An intriguing fact is its
recent observation in the $\omega\phi$ decay channel \cite{Do23}
(see also references herein). Complementary studies of the
$a_0(1710)$ and $f_0(1710)$ resonances are very promising.

\begin{figure}  
\begin{center}\includegraphics[width=16.5cm]{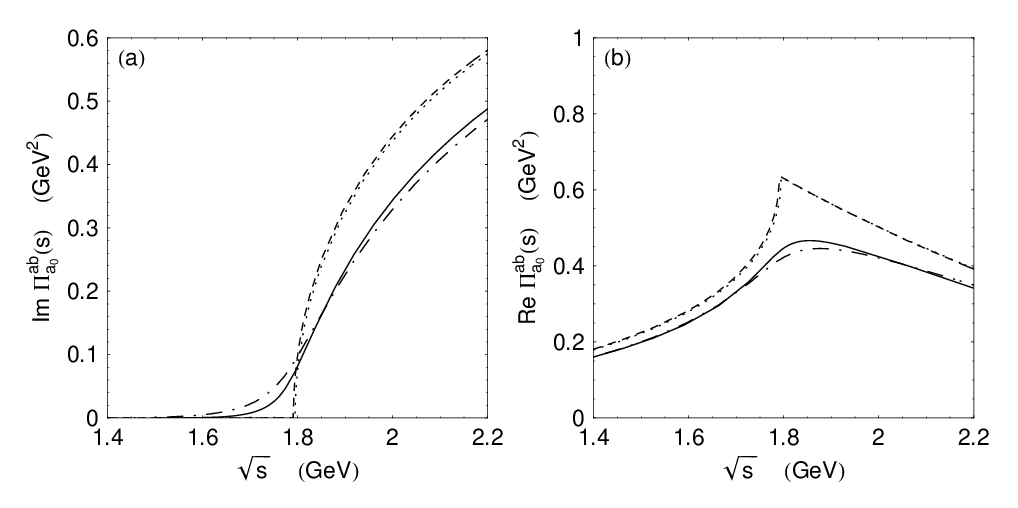}
\caption{\label{Fig1} (a) $\mbox{Im}\,\Pi^{ab }_{a_0}(s)$ and (b)
$\mbox{Re}\,\Pi^{ab}_{a_0}(s)$ as functions of $\sqrt{s}$ for
$ab=K^{*+}\bar K^{*0}$ and $\rho^+\phi$. The dashed and dotted
curves correspond to the $K^{*+}\bar K^{*0}$ and $\rho^+\phi$ decay
channels of the $a_0(1710)$, respectively, for the case of the
stable vector mesons; see Eqs. (\ref{Eq2})$-$(\ref{Eq6}). The solid
and dot-dashed curves correspond to the $K^{*+}\bar K^{*0 }$ and
$\rho^+ \phi$ decay channels, respectively, with taking into account
of the finite widths of the $K^*$ and $\rho$ mesons; see Eqs.
(\ref{Eq9}) and (\ref{Eq12})$-$(\ref{Eq13a}). Here, we put
$g^2_{a_0ab} /(16\pi)=1$ GeV$^2$ to easier understand the scale of
the functions $\mbox{Im} \,\Pi^{ab}_{a_0}(s)$ and
$\mbox{Re}\,\Pi^{ab}_{a_0}(s)$. }\end{center}\end{figure}

Figure \ref{Fig1} shows the imaginary and real parts of the
polarization operators $\Pi^{K^{*+}\bar K^{*0}}_{a_0}(s)$ and
$\Pi^{\rho^+ \phi}_{a_0}(s)$. In the hypothetical case of stable
vector mesons, these functions change sharply near the nominal
$K^{*+}\bar K^{*0}$ and $\rho^+\phi$ thresholds. Accounting for the
finiteness of the widths of the $K^*$ and $\rho$ mesons smooths out
these sharp changes in $\Pi^{K^{*+}\bar K^{*0}}_{a_0}(s)$ and
$\Pi^{\rho^+\phi}_{a_0}(s)$. Physically important here is the
appearance of the quite noticeable energy-dependent widths for the
$a^+_0\to K^{*+} \bar K^{*0}\to(K \pi)^+(\bar K\pi)^0$ and
$a^+_0\to\rho^+ \phi\to \pi^+\pi^0 \phi$ decays for
$\sqrt{s}<2m_{K^*}\approx1.791$ GeV and $\sqrt{s}<m_\rho
+m_\phi\approx1.795$ GeV, respectively. A similar picture takes
place also for $\Pi^{\rho^+\omega}_{a_0}(s)$.

\section{\boldmath $a_0(1710)$ mass spectra}

\begin{figure}  
\begin{center}\includegraphics[width=9cm]{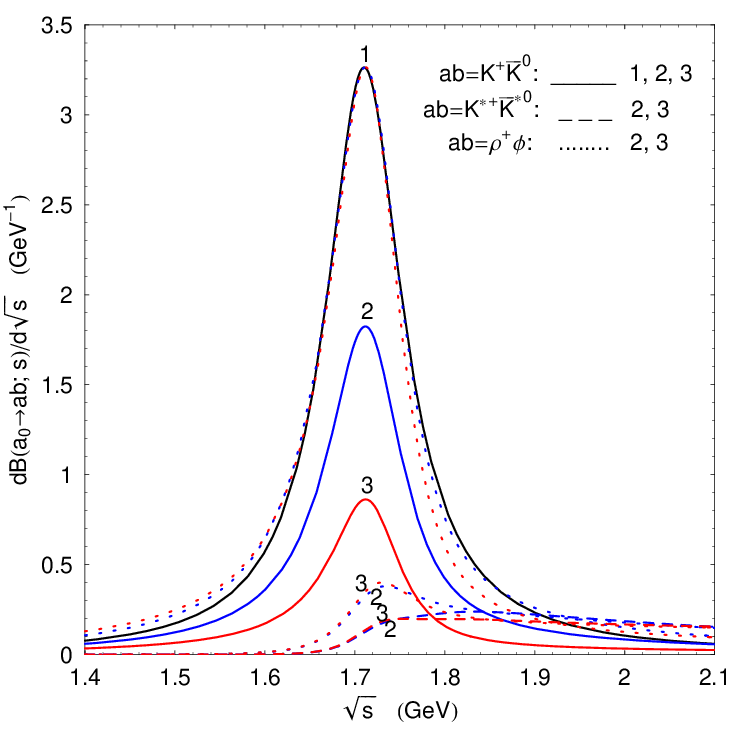}
\caption{\label{Fig2} The mass spectra $d\mathcal{B}(a_0\to ab;\,
s)/d\sqrt{s}$. Solid curves 1$-$3 correspond to the $a^+_0\to
K^+\bar K^0$ decay; dashed curves 2 and 3 to $a^+_0\to K^{*+} \bar
K^{*0}$; dotted curves 2 and 3 to $a^+_0\to\rho^+\phi$. The curve
numbers are attached to the numbers of the value sets of the $a_0$
resonance parameters in Table III. The dotted curves without numbers
which practically coincide with solid curve 1 in the peak region are
explained in the text.}\end{center}\end{figure}

\begin{figure}  
\begin{center}\includegraphics[width=9cm]{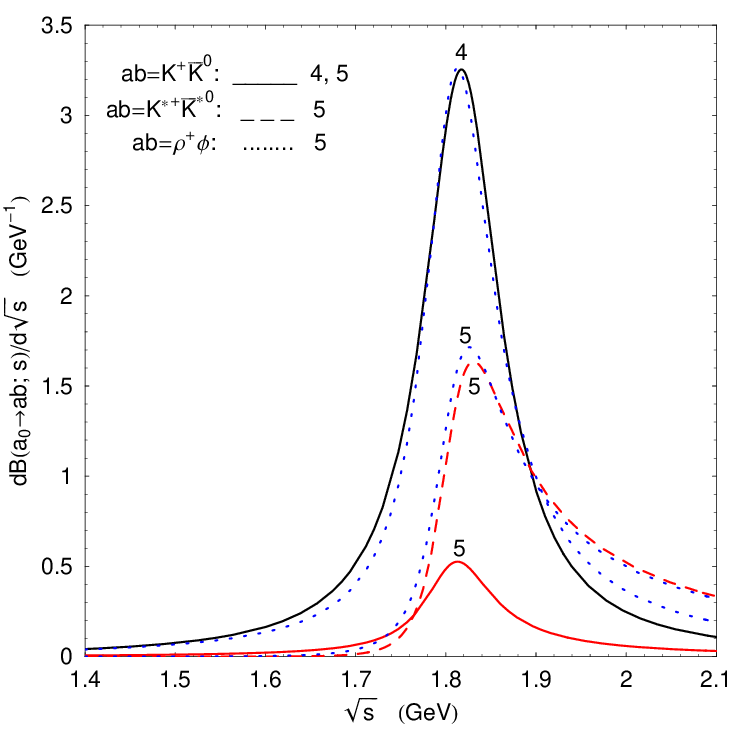}
\caption{\label{Fig3} The mass spectra $d\mathcal{B}(a_0\to ab;\,
s)/d\sqrt{s}$. Solid curves 4 and 5 correspond to the $a^+_0\to
K^+\bar K^0$ decay; dashed curve 5 to $a^+_0\to K^{*+} \bar K^{*0}$;
dotted curve 5 to $a^+_0\to\rho^+\phi$. The curve numbers are
attached to the numbers of the value sets of the $a_0$ resonance
parameters in Table III. The dotted curve without number which
practically coincides with solid curve 4 in the peak region is
explained in the text.}\end{center}\end{figure}

\begin{figure}  
\begin{center}\includegraphics[width=9cm]{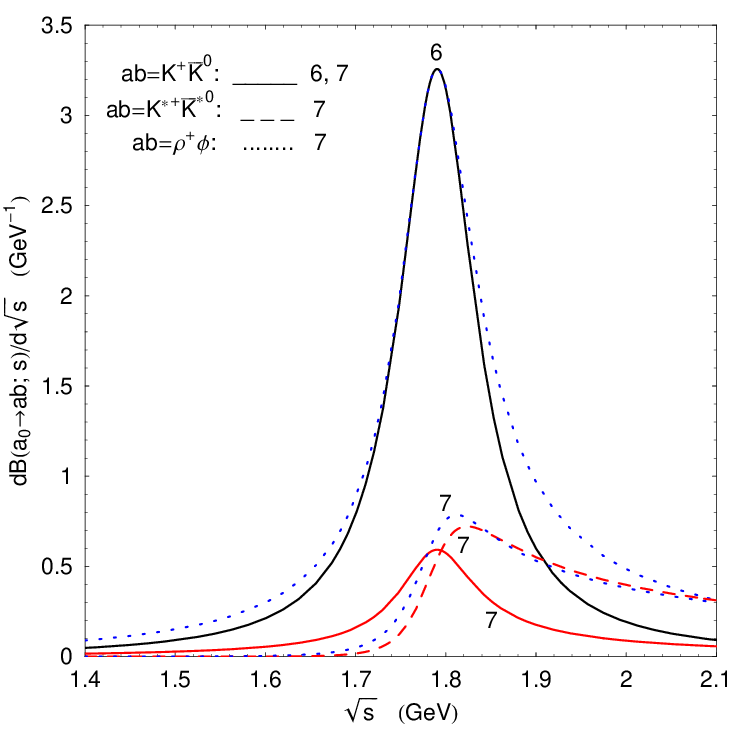}
\caption{\label{Fig4} The mass spectra $d\mathcal{B}(a_0\to ab;\,
s)/d\sqrt{s}$. Solid curves 6 and 7 correspond to the $a^+_0\to
K^+\bar K^0$ decay; dashed curve 7 to $a^+_0\to K^{*+} \bar K^{*0}$;
dotted curve 7 to $a^+_0\to\rho^+\phi$. The curve numbers are
attached to the numbers of the value sets of the $a_0$ resonance
parameters in Table III. The dotted curve without number which
practically coincides with solid curve 6 in the region of the peak
maximum is explained in the text.}\end{center}\end{figure}

Let us write the mass spectrum for the $a_0(1710)\to ab$ decay as
\begin{eqnarray}\label{Eq19} \frac{d\mathcal{B}(a_0\to ab;\, s)}{d\sqrt{s}}=
\frac{2\sqrt{s}}{\pi}\,\frac{\sqrt{s}\Gamma_{a_0\to ab}(s)}{|D_{a_0}
(s)|^2}\end{eqnarray} [we note that in our model $\mathcal{B}(a_0
\to\mbox{all})=1$ in accordance with the unitarity requirement].
Following our first scenario, we assume that the $a_0(1710)$ state
looks like the four-quark state $C^s_\pi(\underline{9}^*)$ and
consider the influence of the coupled $VV$ channels on the shape of
its $K^+\bar K^0$ mass spectrum depending on three parameters
$m_{a_0}$, $g_1\equiv g_{a^+_0K^+\bar K^0}$, and $g_2\equiv g_{a^+_0
K^{*+}\bar K^{*0}}$. In so doing, $g_{a^+_0\eta \pi^+}=-g_1
\cos(\theta_i- \theta_p)$, $g_{a^+_0\eta'\pi^+}=g_1\sin( \theta_i
-\theta_p)$, and $g_{a^+_0\rho^+\phi}=g_2$. Table III presents a few
sets of the values of these parameters which will help us to
understand the general situation. For $m_{a_0}$, it suffices to
consider two extreme values: $m_{a_0}\approx 1700$ MeV and
$m_{a_0}\approx1800$ MeV.
\begin{center}\begin{table}  [!ht]
\caption{Parameters of the $a_0$ resonance. Its mass and widths are
in units of MeV and the coupling constants squared in units of
GeV$^2$. $\Gamma_{a_0\to PP}(m^2_{a_0})$ is the sum of the $a^+_0\to
K^+\bar K^0$, $\pi^+\eta$, and $\pi^+\eta'$ decay widths. For all
variants, the visible width of the $a_0$ peak (i.e., its full width
at half maximum) is about 100 MeV.}
\begin{tabular}{ c c c c c c c } \hline\hline
  \ \,No.\ \ &\ \ $m_{a_0}$\ \ &\ \ $g^2_1/(16\pi)$\ \ &\ $g^2_2/(16\pi)$
  \ \ &\ $g^2_2/g^2_1$\ \ &\ $\Gamma_{a_0\to PP}(m^2_{a_0})$
  \ \ &\ $\Gamma^{\scriptsize\mbox{tot}}_{a_0}(m^2_{a_0})$\ \,\\ \hline
  1 & 1710 & 0.1075  & 0    & 0\ \ &\ 100\ \ &\ 100 \\
  2 & 1717 & 0.15  & 0.84 & 5.6\ \ &\ 139\ \ &\ 159 \\
  3 & 1722 & 0.235  & 3  & 12.8\ \ &\ 218\ \ &\ 234 \\ \hline
  4 & 1817 & 0.1108 & 0  & 0\ \ &\ 100\ \ &\ 100 \\
  5 & 1840 & 0.035   & 0.785 & 22.4\ \ &\ 31\ \ &\ 158 \\ \hline
  6 & 1790 & 0.11 & 0 & 0\ \ &\ 100\ \ &\ 100 \\
  7 & 1830 & 0.11   & 1.25 & 11.4\ \ &\ 99\ \ &\ 281 \\
  \hline\hline\end{tabular}\end{table}\end{center}
Sets 1, 4, and 6 we use as reference. They correspond to the $a_0$
resonance coupled only with pseudoscalar mesons. The values of the
constant $g^2_1/(16\pi)$ have been estimated for these variants
based on the assumption that the total $a_0$ decay width
$\Gamma^{\scriptsize\mbox{tot}}_{a_0}(m^2_{ a_0})=-\mbox{Im}
\,D_{a_0}(m^2_{a_0})/m_{a_0}=100$ MeV [see Eqs. (\ref{Eq1a}) and
(\ref{Eq2})]. The $a_0$ line shapes in the $PP$ decay channels for
variants 1, 4, and 6 correspond to the standard Breit-Wigner
resonance curves with which it is convenient to compare the shapes
of the $PP$ mass spectra corresponding to the cases of the strong
$a_0$ coupling to $VV$ channels. It is obvious that for small values
of the ratio $g^2_2/g^2_1$ the effect of the $a_0$ coupling to
vector mesons on the $PP$  mass spectra is to be small (as well as
the $a_0$ manifestation in $VV$ channels). Therefore, it is
interesting to consider the situation when the ratio $g^2_2/g^2_1$
is much greater than one (as in the $q^2\bar q^2$ model). For its
illustration, we use variants 2, 3, 5, and 7 shown in Table III.
Before proceeding to their analysis, we explain the notation of the
curves in Figs. \ref{Fig2}$-$\ref{Fig4}. For example, the solid
curve labeled by 1 in Fig. \ref{Fig2} represents the mass spectrum
$d\mathcal{B}(a^+_0\to K^+\bar K^0;\, s)/d\sqrt{s}$ corresponding to
the value set of the $a_0$ parameters from Table III heaving the
same number 1. That is, the numbers of the curves in Figs.
\ref{Fig2}$-$\ref{Fig4} are attached to the numbers of the value
sets of the $a_0$ resonance parameters in Table III. Let us start
with a discussion of the mass spectra $d\mathcal{B}(a_0\to ab;\,
s)/d\sqrt{s}$ shown in Fig. \ref{Fig2}. The solid curves 1$-$3 show
the mass spectra $d\mathcal{B} (a^+_0\to K^+\bar
K^0;\,s)/d\sqrt{s}$. The mass spectra in the decays $a_0\to
K^{*+}\bar K^{*0}$ and $\rho^+\phi$ are shown by dashed and dotted
curves 2 and 3, respectively. Naturally, we are talking about the
mass spectra constructed taking into account the instability of the
$K^*$ and $\rho$ mesons. If we multiply solid curves 2 and 3 by 1.79
and 3.785, respectively, and depict results by the dotted curves,
then, as can be seen from the figure, they coincide with a good
accuracy with curve 1 (for which $g_2=0$) in the range $1.62\mbox{
GeV}<\sqrt{s}<1.76\mbox{ GeV}$. Thus, all three $K^+\bar K^0$ mass
spectra have practically the same visible width of the resonance
peaks approximately equal to 100 MeV. At the same time, Table III
indicates that for parameter sets 2 and 3 $\Gamma_{a_0\to
PP}(m^2_{a_0})\approx 139$ and 218 MeV and $\Gamma^{\scriptsize
\mbox{tot}}_{a_0}(m^2_{a_0})\approx159$ and 294 MeV [see Eq.
(\ref{Eq1a})], respectively. The fact that the visible width of the
$a^+_0$ peak in the $K^+\bar K^0$ channel turns out to be noticeably
smaller than $\Gamma_{a_0\to PP}(m^2_{a_0})$ and $\Gamma^{
\scriptsize\mbox{tot}}_{a_0 }(m^2_{ a_0})$ is a direct consequence
of the strong coupling of the $a_0$ with vector channels, which
narrows the $PP$ mass spectra. Deviations of the shape of solid
curves 2 and 3 from curve 1 outside the interval $1.62\mbox{ GeV}
<\sqrt{s}<1.76 \mbox{ GeV}$ are not so large in order that they can
be effectively used to detect the $a_0$ coupling to $VV$ channels.

The above examples corresponding to $m_{a_0}$ at about 1700 MeV show
that the strong coupling of the $a_0$ resonance with the $VV$
channels is undoubtedly possible. At the same time, the shape of the
$K\bar K$ mass spectrum in the region of the $a_0$ peak can be quite
satisfactorily described without taking into account the $a_0 VV$
coupling. Certainly, a similar situation takes place for the decay
channels of the $a_0$ into $\pi\eta$ and $\pi\eta'$. That is, the
strong $a_0$ coupling to $VV$ turns out to be hidden in the $a_0\to
PP$ decay channels, and, therefore one can speak about it only
presumably, at least until direct detection of the decays $a_0\to
VV$.

With $m_{a_0}$ increasing from 1700 to 1800 MeV (and further), the
$a_0\to VV$ decay channels become more and more open and the
contribution from the width $\Gamma_{a_0\to VV}(m^2_{a_0})$ becomes
dominated in the total width $\Gamma^{\scriptsize\mbox{tot}}_{a_0}
(m^2_{a_0})$. Therefore, if we want to retain the visible width of
the $a_0$ peak in the $PP$ channels at a level of about 100 MeV, it
is necessary to reduce the contribution from $\Gamma_{a_0\to PP}(
m^2_{a_0})$ to $\Gamma^{\scriptsize\mbox{tot}}_{ a_0}(m^2_{a_0})$
[i.e., decrease $g^2_1/(16\pi)$]. Let us illustrate the above by
using variants 4 and 5 in Table III in which $m_{a_0}$ takes the
values slightly above the nominal thresholds of the $K^{*+}\bar
K^{*0}$ and $\rho^+\phi$ decay channels. The corresponding mass
spectra are shown in Fig. \ref{Fig3}. The $K^+\bar K^0$ mass spectra
are shown by solid curves 4 and 5. Dashed curve 5 and dotted curve 5
show the mass spectra in the $a_0$ decay channels into $K^{*+}\bar
K^{*0}$ and $\rho^+\phi$, respectively. It is worth to pay attention
to the difference between the mass distributions of $K^{*+}\bar
K^{*0}$ and $\rho^+\phi$ from the case of $K^+\bar K^0$. If the
solid curve 5 is enlarged 6.2 times and depicted as a dotted one,
then, as can be seen from the figure, its shape repeats the shape of
curve 1 [for which $g_2=0$ and $\Gamma^{\scriptsize \mbox{tot}
}_{a_0}(m^2_{a_0}) =100$ MeV] with a good accuracy in the interval
$1.76\mbox{ GeV}< \sqrt{s}<1.9\mbox{ GeV}$. Note that solid curve 5
corresponds to $\Gamma^{\scriptsize\mbox{tot}}_{a_0} (m^2_{a_0})
\approx158$ MeV and $\Gamma_{ a_0\to PP}(m^2_{a_0})\approx31$ MeV
(see Table III). Thus, for $m_{a_0}\approx1800$ MeV, the strong
$a_0$ coupling to $VV$ pairs remains hidden as before, i.e., has a
very little effect on the visible width and shape of the mass
distributions in the $PP$ channels. On the other hand, the values of
$m_{a_0}\approx1800$ MeV favor the bright manifestation of the $a_0$
resonance in the $K^{*+}\bar K^{*0}$ and $\rho^+\phi$ mass spectra,
as can be seen from the comparison of the corresponding curves in
Figs. \ref{Fig2} and \ref{Fig3}.

Let now the constant $g^2_1/(16\pi)$ takes the values greater than
in variant 5 indicated in Table III. The corresponding mass spectra
are shown in Fig. \ref{Fig4}. They correspond to variants 6 and 7 in
Table III. If the solid curve 7 is magnified by a factor of 5.5 and
depicted as a dotted one, then, as can be seen from the figure, its
right wing deviates noticeably from the reference curve 6 [for which
$g_2=0$ and $\Gamma^{\scriptsize\mbox{tot}}_{a_0}(m^2_{a_0} )=100$
MeV]. Such a shape asymmetry of the $K^+\bar K^0$ mass spectrum can
be discovered, in principle, provided that the background
contributions accompanying the $a_0$ resonance are small and the
region of high $\sqrt{s}$ is not limited by the phase space of the
reaction.

\begin{figure}  
\begin{center}\includegraphics[width=9cm]{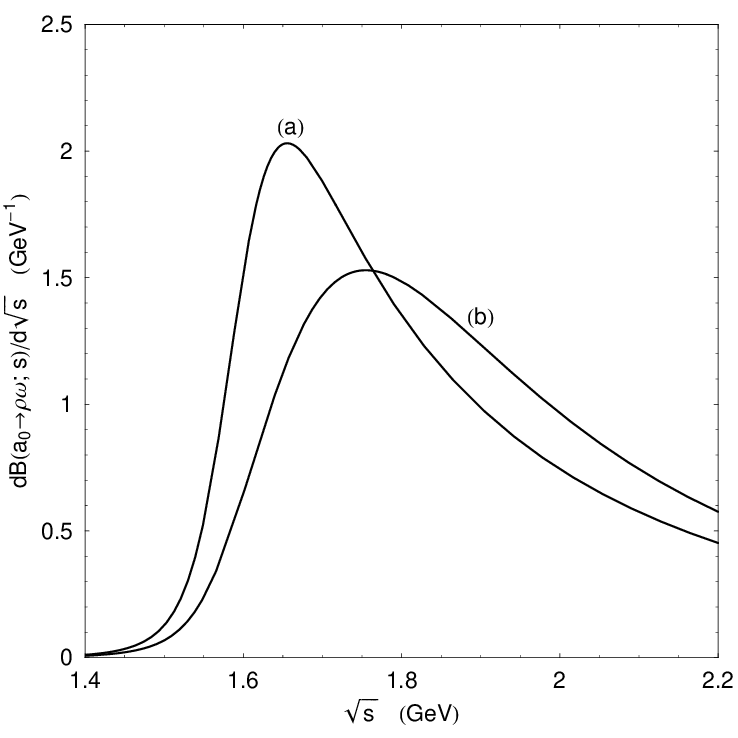}
\caption{\label{Fig5} Curves (a) and (b) show the mass spectrum
$d\mathcal{B}(a_0\to\rho \omega;\,s)/d\sqrt{s}$ calculated at
$m_{a_0}=1710$ and 1817 MeV, respectively. See the text for
details.}\end{center}\end{figure}

Finally, consider the scenario where the $a_0$ state is similar to
the $q^2\bar q^2$ state $C_\pi(\underline{36}^*)$; see (\ref{Eq15}).
The coupled channels in this case are $\pi\eta$, $\pi\eta'$, and
$\rho\omega$ channels; the $a_0$ coupling to the latter is dominant;
see Table II. The $\rho\omega$ channel is the open one, since its
the nominal threshold is approximately equal to 1558 MeV, and the
$a_0$ resonance decaying into $\rho\omega$ is in the region of
1700$-$1800 MeV. Adhering to the $q^2\bar q^2$ model, we can express
the coupling constant $a_0$ to $\rho\omega$ in terms of the constant
$g_2$ introduced above: $g_3\equiv g_{a_0\rho\omega}=0.743g_0=
-\sqrt{2}(0.743/0.644) g_2\approx-1.63g_2$ [see Eqs. (\ref{Eq14})
and (\ref{Eq15}) and Table II]. Since $g^2_3\approx2.66g^2_2$, the
$a_0$ resonance can be very broad. As an illustration, we set
$g^2_2/(16 \pi)=0.75$ GeV$^2$ [for comparison see the values of
$g^2_2/(16\pi)$ indicated in Table III]. We also neglect a tiny
$a_0$ coupling to $\pi\eta$ and $\pi\eta'$ channels [see Eq.
(\ref{Eq15})]. The corresponding examples of the mass spectrum
$d\mathcal{B}(a_0\to \rho\omega;\,s)/d\sqrt{s}$ are shown in Fig.
\ref{Fig5}. The curves (a) and (b) correspond to
$d\mathcal{B}(a_0\to\rho\omega;\,s)/d \sqrt{s}$ calculated at
$m_{a_0}=1710$ and 1817 MeV, respectively. Note that the total
widths of distributions (a) and (b) calculated by Eq. (\ref{Eq1a})
at $\sqrt{s}=m_{a_0}$ are of about $600$ and $800$ MeV,
respectively. At the same time, their visible widths turn out to be
approximately 300 and 500 MeV, respectively, owing to the
energy-dependent finite width corrections in the $a_0$ propagator;
see (\ref{Eq1}).

Thus, the $a_0$ resonance which is discovered in the $K\bar K$ and
$\pi\eta$ decay channels in the region of 1700$-$1800 MeV  can be a
manifestation of the four-quark  state $C^s_\pi(\underline{9}^*)$.
This state can also manifest itself in the $K^*\bar K^*$ and  $\rho
\phi$ decay channels that have not yet been studied. In addition, a
very broad  $\rho\omega$ enhancement at about 1600$-$1800 MeV (if
found) can be due to the $C_\pi(\underline{36}^*)$ state. In Ref.
\cite{FN}, for completeness, we provide a brief explanation of our
choice of all above variants.

\section{\boldmath Conclusion}

The discovery of a new rather narrow, heavy, isovector scalar meson
$a_0(1700/1800)$ decaying to $K\bar K$ and $\eta\pi$  was a
sufficiently unexpected event \cite{Ab21,Le21,Ab22a,Ab22b,Aa23}.
From the theoretical considerations mentioned in the introduction,
it follows that $a_0(1710)$ can be strongly coupled to decay
channels into two vector mesons. Note that good probes for the
search for the $a_0\to VV$ decays can be reactions in which the
$a_0$ state was observed in the $PP$ decay channels: $D^+_s\to
K^+K^-\pi^+$ \cite{Ab21}, $\gamma\gamma\to\eta_c \to\eta\pi^+\pi^-$
\cite{Le21}, $D^+_s\to K^0_SK^0_S\pi^+$ \cite{Ab22a}, $D^+_s\to
K^0_SK^+\pi^0$ \cite{Ab22b}, and $B^+\to [\eta_c,\,\eta_c(2S),\,
\chi_{c1}]K^+\to(K^0_SK^\pm\pi^\mp)K^+$ \cite{Aa23}. A discussion of
the processes that have the potential for detecting $a_0(1710)\to
PP,VV$ decays can also be found in Refs. \cite{Os23b,Wan23,Di23}. Of
course, the detection of the decays $a_0\to K^*\bar K^*$,
$\rho\phi$, $\rho \omega$ is not an easy task in all the $a_0$
production reactions. For example, in the processes mentioned above,
it will be necessary to study the $K\bar K\pi\pi\pi$ or $6\pi$ final
states (instead of $K\bar K\pi$ and $\eta\pi\pi$) in order to
extract the contributions of the quasi-two-body components
$\rho\phi$, $K^*\bar K^*$, or $\rho\omega$ within the isobar model.
In this paper, we did not set ourselves the goal of studying
specific processes involving the production of the vector meson
pairs in final states. Each such process is unique and requires
special consideration. Nevertheless, we hope that our work
contributes future investigations just in this direction. The mass
spectra constructed by us for the  decays $a_0\to K^{*+}\bar
K^{*0}$, $a_0\to\rho^+\phi$, and $a_0\to\rho^+\omega$ in Figs. 2$-$5
could be confronted with future data.

In this paper, we have analyzed the effect of the strong coupling of
the $a_0(1710)$ resonance to vector-vector channels on its line
shape in the decay channels into two pseudoscalar mesons. We have
assumed that the $a_0(1710)$ state strongly coupled to the $VV$
channels might be similar to the four-quark state belonging to
either the $\underline{9}^*$ or $\underline{36 }^*$ multiplet. Our
goal was to find evidence in favor of the strong coupling of the
$a_0(1710)$ to vector mesons in the decay channels into pseudoscalar
mesons. The impetus to this was the well-known effect of the
narrowing of the $\pi\eta$ mass spectrum in the $a_0(980)\to\pi
\eta$ decay caused by the influence of the strong coupling
$a_0(980)$ to the $K\bar K$ channel \cite{Fl76,Ac80a}. This effect
helped, in particular, to eliminate the obvious contradiction
between the observed narrowness of the $a_0(980)$ peak in the
$\pi\eta $ channel and the assumption of the $q^2\bar q^2$ model
about the superallowed coupling of the $a_0(980)$ to $\pi\eta$ and
$K\bar K$ channels \cite{Ac80a, Ac80b}. We have found out that in
the case of the state $a_0(1710)$ its strong coupling to $VV$
channels can work similarly to the coupling of the $a_0(980)$ to
$K\bar K$, i.e., to narrow the $a_0(1710)$ peak in the $PP$ mass
spectra. If, in the presence of this coupling, the visible width of
the $a_0(1710)$ in the $K\bar K$ and $\pi\eta$ channels turns out to
be (as in experiment) about 100 MeV, then, in its absence, the width
could be from 150 to 300 MeV. At present, the fundamental difference
between the situation with the $a_0(1710)$ and the situation with
the $a_0(9 80)$ is in the absence of the data on $a_0(1710)\to VV$
decays. In addition, we have shown that in many cases the shape of
the $a_0(1710)$ mass spectra in $PP$ channels can be satisfactorily
described without taking into account the $a_0(1710)VV$ coupling at
all. That is, the $a_0(1710)$ coupling to $VV$ turns out to be
hidden in the $PP$ channels, and so far it is possible to speak
about its existence only hypothetically. Additional investigations
are needed in this direction, and first of all, of course, the
direct detection of the $a_0(1710)\to VV$ decays is necessary.
Combined studies of the $a_0(1710)\to VV$ and $f_0(1710)\to VV$
decays are also very promising.\\

The work was carried out within the framework of the state contract
of the Sobolev Institute of Mathematics, Project No. FWNF-2022-0021.


\begin{thebibliography}{99}
\bibitem{FN} At present, the position of $m_{a_0}$ is not well defined. In
Sec. II, we noted that in the $q^2\bar q^2$ model the values of
$m_{a_0}$ in the interval 1700$-$1800 MeV are quite admissible. In
the literature, the values $m_{a_0}=1710$ and 1817 MeV are the most
frequently mentioned (tentative) values. Therefore, we used these
values for $m_{a_0}$ in reference sets 1 and 4 in Table III. Set 6
with $m_{a_0}=1790$ MeV from the considered interval we also used as
the reference one. Parameters $a_0$ for sets 2, 3, 5, and 7 were
selected so as to best reproduce the mass spectra for the
corresponding reference variants (for which $g_2=0$). In so doing,
all the given examples had to correspond to large values of the
ratio $g^2_2/g^2_1$ (as indicated by the $q^2\bar q^2$ model), as
well as the visible width of the $a_0$ peak in the $PP$ channels of
about 100 MeV. It is clear that the selected examples for the
$a_0=C^s_\pi(\underline{9}^*)$ scenario are quite sufficient to
understand the possible situation in the $a_0$ mass range from 1700
to 1800 MeV. The choice of the $a_0$ parameters to illustrate the
$a_0=C_\pi(\underline{36}^*)$ scenario is exhaustively described in
the text.
\end{thebibliography}
\end{document}